\documentclass[onecolumn,secnumarabic,amssymb, nobibnotes, aps,notitlepage, pr,superscriptaddress,10pt]{revtex4-2}

\usepackage{amsmath,amssymb}
\usepackage{stackengine,graphicx}
    \graphicspath{{images/}} 
\usepackage[dvipsnames]{xcolor}
\usepackage{bbm}
\usepackage{booktabs}
\usepackage{subcaption}
\usepackage{enumitem}
\usepackage{colortbl}
\usepackage{lipsum}

\usepackage{tabu}
\usepackage{makecell}

\usepackage{xurl}

\PassOptionsToPackage{hyphens}{url}\usepackage{hyperref}

\usepackage{subcaption}

\usepackage{algorithm}
\usepackage{algpseudocode}
\usepackage{soul}

\usepackage{collectbox}
\usepackage{bbm}
\usepackage{capt-of}

\UseRawInputEncoding
\usepackage[utf8]{inputenc}

\usepackage{hyperref}
\hypersetup{
    colorlinks=true,
    linkcolor=blue,
    filecolor=magenta,      
    urlcolor=cyan,
    citecolor=black
}

\usepackage{array}
\newcolumntype{L}[1]{>{\raggedright\let\newline\\\arraybackslash\hspace{0pt}}m{#1}}
\newcolumntype{C}[1]{>{\centering\let\newline\\\arraybackslash\hspace{0pt}}m{#1}}
\newcolumntype{R}[1]{>{\raggedleft\let\newline\\\arraybackslash\hspace{0pt}}m{#1}}

\usepackage{accents}

\newcommand{\lu}{College of Health, Lehigh University, Bethlehem, Pennsylvania, United States of America}

\newcommand{\cee}{Department of Civil and Environmental Engineering, P.C. Rossin College of Engineering and Applied Science, Lehigh University, Bethlehem, Pennsylvania, United States of America}

\newcommand{\metaculus}{Metaculus, Santa Cruz, California, United States of America}

\newcommand{\gjo}{Good Judgment Inc., New York, New York, United States of America}

\newcommand{\columbia}{Department of Epidemiology, Mailman School of Public Health, Columbia University, New York, United States of America}

\newcommand{\MIT}{Massachusetts Institute of Technology, Cambridge, Massachusetts, United States of America}

\date{\today} 

\usepackage[paperwidth=8.5in,paperheight=11.0in,
  left=1.in,right=1.in,top=1.5in,bottom=1.5in,
  includefoot,heightrounded]{geometry}

\begin{document}

  \title{Aggregating human judgment probabilistic predictions of COVID-19 transmission, burden, and preventative measures}

  \author{Allison~Codi}
  \affiliation{\lu}
  
  \author{Damon~Luk}
  \affiliation{\lu}
  
  \author{David~Braun}
  \affiliation{\lu}
  
  \author{Juan~Cambeiro}
  \affiliation{\metaculus}
  \affiliation{\columbia}
  
  \author{Tamay~Besiroglu}
  \affiliation{\metaculus}
  \affiliation{\MIT}
  
  \author{Eva Chen}
  \affiliation{\gjo}
  
  \author{Luis Enrique Urtubey de C\`{e}saris}
  \affiliation{\gjo}
  
  \author{Paolo~Bocchini}
  \affiliation{\cee}
  
  \author{Thomas~McAndrew}
  \email{mcandrew@lehigh.edu}
  \affiliation{\lu}

  \setlength{\parskip}{0.5em}
  
  \setlength{\parindent}{0pt}
  
  \begin{abstract}
  
  \begin{flushleft}
      \textbf{Abstract:} 
      Aggregated human judgment forecasts for COVID-19 targets of public health importance are accurate, often outperforming computational models.
      Our work shows aggregated human judgment forecasts for infectious agents are timely, accurate, and adaptable, and can be used as tool to aid public health decision making during outbreaks.
  \end{flushleft}
      
  \end{abstract}

  \maketitle
  
  \clearpage
  
  \section{Introduction}
  
  Accurate forecasts of the trajectory of COVID-19 and preventative measures to reduce transmission of SARS-CoV-2 provide foresight that enables public health officials to mitigate the impact of the pandemic~\cite{pollett2021recommended}.  
  Mathematical models are the most often used tool to improve situational awareness~\cite{biggerstaff2021improving}. 
  However, most mathematical models rely on structured, reported surveillance data and often do not have access to community level transmission dynamics, data related to human behavior, or behavioral responses to policy changes.
  
  Human judgment has produced accurate forecasts of the evolution of an infectious agent for seasonal epidemics and pandemic events~\cite{farrow2017human,mcandrew2020expert}. 
  Past work studying COVID-19 and human judgment has highlighted the ability of aggregate human judgment predictions to adapt to changing dynamics faster than mathematical models~\cite{bosse2021comparing}. 
  When human judgment forecasts have had lower accuracy than mathematical models, previous work has shown that combining the two improves performance over the mathematical model alone~\cite{ibrahim2021}. 
  Human judgment predictions of an infectious agent are low-overhead, flexible, and supply rapid and adaptable forecasts to public health decision makers ~\cite{mcandrew2020expert}. 
  
  To best prepare for and prevent infectious disease outbreaks, health officials need quick, accurate, and adaptable forecasts~\cite{lutz2019applying}.
  We show evidence that supports human judgment aggregated probabilistic predictions meet these criteria for COVID-19 targets associated with transmission, burden, and preventative measures.
  
  \section{Methods}
  
  Monthly surveys from Jan.\ 6, 2021 to Jun.\ 16, 2021 collected predictions from two human judgment forecasting platforms: Metaculus and Good Judgment Open (GJO)~\cite{metaculus,gjo}.
  Subscribers to both platforms were invited to participate via email solicitation.  
  We included monthly forecasts of the pandemic in summary reports to aid real-time public health decision-making which contain a detailed list of human judgment predictions and the exact wording of each question posed to both crowds.~\cite{summaryReports}.
  
  Participants had approximately twelve days to provide probabilistic predictions one to three weeks ahead of time at the US national level for six targets of public health importance:~(1) weekly incident cases, (2) hospitalizations, (3) deaths, (4) cumulative first and (5) full-dose vaccinations, and (6) prevalence of immunity evading variants.
  Participants could submit an initial prediction and revise their prediction as many times as they wished within the twelve-day period. 
  Participants received feedback about the accuracy of their forecast via email when the ground truth was available.
  
  Individual forecasts submitted to Metaculus and GJO forecasting platforms were combined into an equally weighted linear pool called a consensus forecast.
  
  Consensus forecasts of incident cases and deaths were compared to the COVID-19 Forecasthub, an ensemble that combined up to 48 computational models between the months of Jan.\ 2021 and Jun.\ 2021 ~\cite{Cramer2021-hub-dataset}.
  The date that forecasts were generated by human judgment and by computational models in the COVID-19 Forecasthub were chosen to be on average within 2 days of one another.
  
  For each target, we report the absolute error~(AE), defined as a forecast median prediction minus the truth, and the percent error~(PE) defined as the absolute error divided by the truth and multiplied by 100. 
  
  \section{Results}
  
  A total of 404 unique participants (71 Metaculus, 333 GJO) submitted probabilistic predictions across the 33 questions for the above six targets for a total of 2,021 unique forecasts~(open access data set available here~\cite{zoltarArc}).
  A participant was not required to answer all questions.
  The median consensus prediction for targets 1-5 had a mean PE of 39\% in the first survey, 9\% for survey 2, 13\% for survey 3, and 11\%, 26\%, 9\% for surveys 4 through 6. The largest PE was 73\% for a prediction of incident cases that was submitted on survey 5 and smallest PE was 0.1\% for a prediction of incident deaths that was submitted on survey 1.
  
  PE for the majority of targets decreased over time. 
  The PE of the median consensus prediction was 58\% (620,192 AE) for incident cases and 60\% (49,201 AE) for incident hospitalizations in the first survey.
  Both targets reduced their PE to 15\% (An AE of 13,803 for cases and an AE of 2,191 for hospitalizations) in the last survey.
  PE decreased from 18\% to 2\% (9,613,628 AE to 3,821,920 AE) for cumulative first-dose vaccinations and from 6.1\% to 5.8\% (3,745,157 AE to 9,236,130 AE) for cumulative full vaccinations between the initial and final surveys.
  
  The PE for median consensus predictions of incident deaths was on average 7\% (451 mean AE across all six surveys) with a PE less than 0.5\% for survey 1 and survey 4 (27 AE and 13 AE).
  
  The PE for variant prevalence was on average 57\% (13 average AE) and the highest PE was 153\% (14 AE) in survey 6.
  
  The median consensus prediction was closer to the truth than 62\% of the 2,021 individual predictions.
  When subset to the six incident deaths targets, the consensus prediction was closer to the truth than 75\% of individual predictions and in survey five the consensus median prediction of incident deaths was closer to the truth than all of the fifty-nine individual predictions. 
  
  Compared to ensemble predictions made by the COVID-19 Forecasthub, the median consensus prediction generated by humans was closer to the truth for 3/6 predictions of incident cases and 4/6 predictions of incident deaths.
  For predictions of incident cases, the mean PE was 32.8\% for the COVID-19 Forecasthub and 33.5\% for aggregate human judgment.
  For incident deaths, the mean PE was 10\% for the COVID-19 Forecasthub vs 7\% for human judgment. 
  
  \section{Discussion}
  
  We show that (i) aggregate human judgment forecasts are frequently closer to the truth than individual forecasts, (ii) the accuracy of aggregate forecasts depends on the target, (iii) the accuracy of aggregate forecasts can improve over time, and (iv) aggregate human judgment can produce forecasts of incident cases and deaths with similar accuracy to an ensemble of computational models.
  
  We are limited by the small number of questions we asked, the short time span over which we surveyed the crowd, and the lack of a controlled environment in which to pose questions.
  
  Contrary to recent work that showed a crowd can produce more accurate forecasts for cases than deaths~\cite{bosse2021comparing}, we found aggregate median predictions of incident deaths were more accurate than predictions of incident cases.
  This may be because humans have the innate capacity to learn relationships between a set of evolving signals, such as incident cases, hospitalizations, and vaccinations, that are correlated with the target they aim to predict. 
  The lack of signals and environmental cues related to questions about the prevalence of specific variants may be why these aggregate forecasts were inaccurate.
  The availability of environmental cues related to cases, deaths, and hospitalizations may explain why participants were able to learn over time, however more experimental work related to how humans incorporate data to make predictions should be explored.
  
  \begin{figure}[ht!]
      \centering
      \includegraphics[scale=0.90]{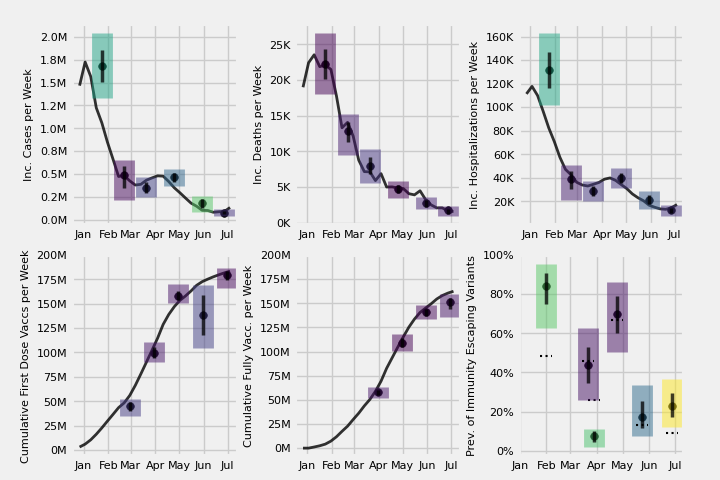}
      \caption{Consensus median~(black dot), 25th and 75th percentiles~(bottom and top of solid black bar), and the 2.5th and 97.5th~(bottom and top of rectangle) for 1 through 3 week ahead predictive distributions of aggregate human judgment forecasts of weekly incident cases, hospitalizations, and deaths, cumulative first and full-dose vaccinations, and prevalence of immunity evading variants at the US national level. Predictions were submitted between Jan.\ 2021 and Jun.\ 2021. Predictions for survey 6 were made for the week starting on Jun.\ 27 and ending on Jul.\ 3. The ground truth is a solid black line or a dashed black line. Lighter rectangles correspond to higher percent error. \label{fig.hjepiweek}}
  \end{figure}
  
  \section{Acknowledgements}
  
  This research was supported through the MIDAS Coordination Center (MIDASNI2020-1) by a grant from the National Institute of General Medical Science (3U24GM132013-02S2). 
  We wish to thank Phillip Rescober for data science support from Good Judgment Inc. 
  We wish to thank all of the individual forecasters who contributed their time and energy to generate predictions about the trajectory of COVID-19.
  
  \clearpage

\end{document}